\documentclass[preprint]{acm_proc_article-sp}
\usepackage{epsfig}
\usepackage[square,comma,numbers,sort&compress]{natbib}
\usepackage{multirow}

\newcommand{\UPLB}{University of the Philippines Los Ba\~{n}os}
\begin{document}

\title{On Web-grid Implementation Using\\ Single System Image}
\numberofauthors{1}
\author{
\alignauthor Marie Yvette B. de~Robles, Zenith O. Arnejo and Jaderick P. Pabico\\
   \affaddr{Institute of Computer Science}\\
   \affaddr{College of Arts and Sciences}\\
   \affaddr{\UPLB}\\
   \affaddr{College 4031, Laguna, Philippines}\\
   \affaddr{63-49-536-2313}\\
   \email{\{mybderobles,zoarnejo,jppabico\}@uplb.edu.ph}
}
\date{}
\toappearbox{6th Symposium on the Mathematical Aspects for Computer Science (SMACS 2012), Boracay Island, Aklan, 04-08 December 2012}

\maketitle

\begin{abstract}
With the latest innovations and trend towards personalizing users' web browsing experience, the web has been increasingly dominated by dynamic contents. However, delivering dynamic content remains a challenge due to the many dependencies involved in compiling the content, specifically personalized ones. This paper presents the use of Single System Image (SSI) clustering systems for a cheap, off-the-shelf, local lightweight distributed web-grid composed of desktop PCs. The three clustering systems considered in the study are Kerrighed, OpenSSI and openMosix. Through an online simulation technique, the performance savings achieved by the clustering systems were measured. Results showed that Kerrighed has the least number of missed requests while the response time is comparable with the rest.
\end{abstract}


\section{Introduction}
Web architectures today suffer degradation and disruptions due to changing workloads caused by 1) time-of-day (TOD) effects; a diurnal variation in traffic on most Web sites observed throughout the day, and 2) flash crowd arrivals (FCA); an unexpected magnitude of users accessing a specific website at an unexpected time. As a result, web architectures are modified to adapt to peak workloads which results to a significant decrease in resource utilization. Hence, there is a need for web architectures allocating the optimal number of needed servers for active workloads. However, the challenge in designing an optimal resource allocation policy is to minimize server resources without affecting the user-perceived Quality of Service (QoS).

To reduce network latency among client and data, companies such as Google and Yahoo utilizes Content Delivery Networks (CDNs) alongside mirroring and caching strategies. Using distributed server architecture, clients are redirected to spatially closest server to provide both scalability and transparency. However, distributed server architectures are specially designed to improve QoS of static content that stands in contrast with the latest trend on web becoming increasingly dominated by dynamic content. Static and dynamic content have different demands on required server resources. Also, dynamic content, being computationally intensive, requires server selection mechanism that may not be optimal for static contents.

Thus, a number of researches has been conducted to improve the performance of both static and dynamic content server policies on web architectures.  Static server policies focus on reduction of end-to-end network latency thus redirecting requests to the nearest server. With this policy, architectures implement either: 1)~shared server utility model wherein a server may process more than one service at  a time; or 2)~full server utility model where each server process one service at a time. On the other hand, dynamic content server policies are proposed to perform caching at either the client or server side as one of the best solutions for server selection.

\begin{figure}[htb]
        \centering
        \epsfig{file=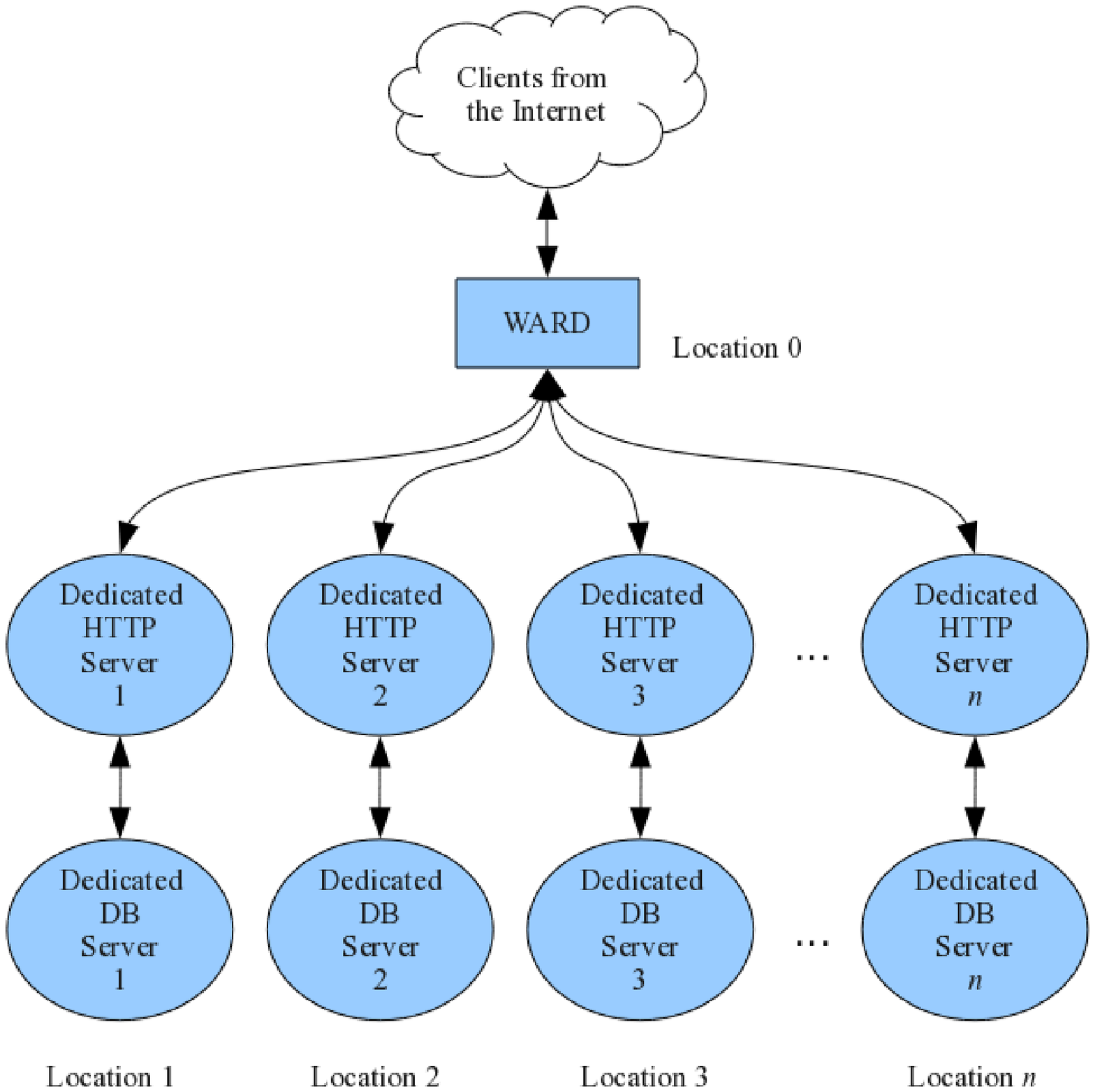, width=2.8in}
        \caption{The conceptual framework of a WARD architecture. Dedicated HTTP and DB servers maybe geographically located differently. This figure is in color in the electronic copy of this paper printed with permission from~\citet{derobles11} and the Computing Society of the Philippines.}
        \label{ward}
\end{figure}

There are also different redirection algorithms proposed to address both static and dynamic content server policies. However, most of it focused on client-side mechanisms that considers network as the primary bottleneck. Recent developments showed that due to the increase of dynamic content web sites and the existence of high interconnection links, network latency is not the main issue anymore, but internal processing brought about by dynamic content web sites. Today, there are few redirection algorithms which tries to solve this problem. One of which is the wide-area redirection (WARD) algorithm where requests are redirected to the best server, either geographically located locally or globally (see Figure~\ref{ward}). The purpose of WARD is to minimize the total networking and server processing delays.  However, WARD may only apply to large data centers that cater millions of requests per day, such as Google, and to those hosting architecture comprising of clusters that are mostly located remotely. A typical web host may not need too many servers scattered globally, but other resources which may be already available locally.  Further, WARD failed to consider the same demand on static-content websites.

\begin{figure}[htb]
        \centering
        \epsfig{file=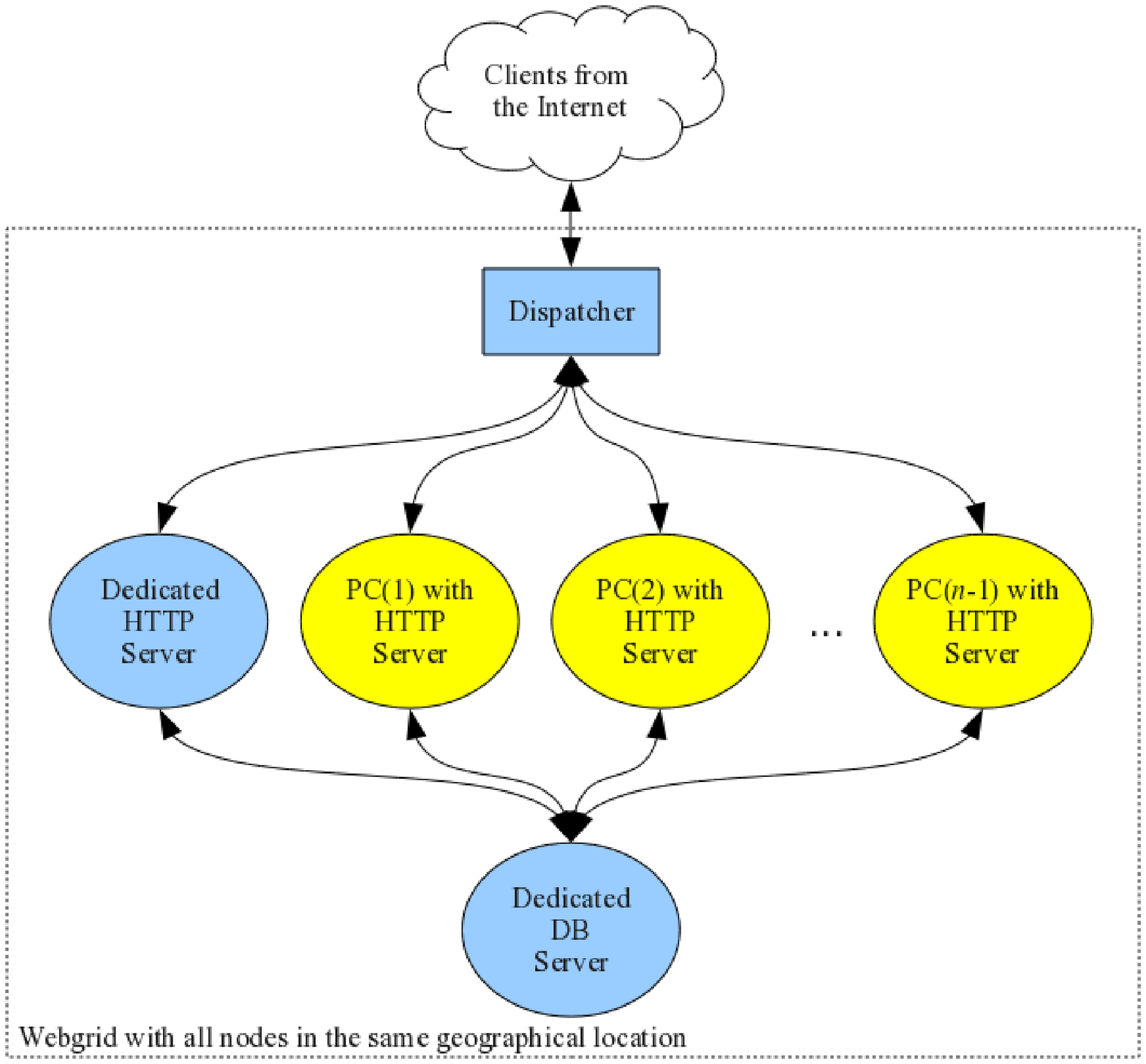, width=2.8in}
        \caption{The conceptual framework of a WARD architecture. Dedicated HTTP and DB servers maybe geographically located differently. This figure is in color in the electronic copy of this paper printed with permission from~\citet{derobles11} and the Computing Society of the Philippines.}
        \label{webgrid}
\end{figure}

A study has been conducted to create a low-cost, off-the-shelf architecture for static and dynamic-content web sites ~\citep{derobles11}. In this work, the elements of the desktop web-grid are personal computers commonly used in simple tasks. These pc's are connected locally to a server and is used to observe outcomes of the Local Area Redirection (LARD) per-request (or per-query) redirection policies (see Figure~\ref{webgrid}).

Results showed that it was able to improve the performance of both static and dynamic-content Web sites, characterized by client access time and resource utilization, even during overload conditions. However, only the Round Robin (RR) and Least Busy Server First (LBSF) algorithms were implemented. Thus, this study proposes the use of Single System Image (SSI) in a desktop web-grid. The three SSI clustering systems considered in the study were Kerrighed, OpenSSI and openMosix.

\section{Review of Literature}
Content Delivery Networks (CDNs)~\citep{presti02,lu08,leong09}, mirroring~\citep{myers99,gavrilovska01}, and caching~\citep{zou03,sosasosa03,bhattacharjee05} are some of the most widely used strategies to improve the performance of static server allocation policies. In CDNs, servers are deployed in multiple data centers and user requests are redirected to the closest server. Caching, on the other hand, keeps frequently accessed information in a location close to the requester. Caching can occur either at the client side or at the server side. At the client side, expiration times are set using cookies while at the server side, cached pages expire upon receiving database update queries. While, mirroring deploys mirror sites across different geographical locations, that is a complete replicate of the origin server. If the primary server fails, the backup server can immediately take its place. These three techniques considered network as the primary problem. However, with the increasing trend towards dynamic content web sites, these mechanisms may not be applicable. In some cases, it may make sense to redirect a dynamic-content request to the server that is geographically furthest if it has the lowest sum of network latency and expected server processing time.

Thus, recent developments focused on the internal complexity brought about by dynamic content web sites. Wide-area Redirection (WARD) mechanism was implemented in ~\citet{ranjan08}, where requests are redirected to the server with the minimum total networking and processing delays. However, these architectures comprise mainly of servers as elements, which could either be hosting a specific or variable contents.  These architectures were also designed for large data centers located globally.  Nonetheless, these kinds of architectures failed to consider the increasing complexity of static content web sites.

In 2011, ~\citet{derobles11} conducted a study on Local Area Redirection (LARD) per-request redirection algorithms . The study observed the performance of LARD given a number of personal computers connected locally to the server as elements in the web-grid. The results, characterized by client access time and resource utilization even during overload conditions, prove that LARD was able to improve the performance of both static and dynamic-content web sites. Utilization of resources, especially servers or computers were also optimized. However, only two redirection algorithms were implemented for the architecture (RR and LBSF). Thus, this study proposes the use of SSI in a desktop web-grid. The SSI considered in the study were Kerrighed, OpenSSI and openMosix.

There was a comparative study of Kerrighed, openMosix and OpenSSI which evaluated the performance of some SSI features using two nodes ~\citep{lottiaux04}. The study was applied to simple sequential application (vector addition) to evaluate process migration and NetPipe for sockets (INET and Unix) and pipes to evaluate stream migration. Since the study by Lottiaux et. al was not able to observe the behavior of the three SSI's given increasing number of nodes and if whether the SSI was subjected to web applications, another contribution of this study is on evaluating the effect of increasing workload and having simple and complex computations and queries.

\subsection{Single Sytem Image}
A single system image (SSI) is the property of a system that makes a cluster of machines appear to be a single system to users. Furthermore, SSI is fault tolerant and is able to manage and schedule resources ensuring that the system is evenly loaded ~\citep{buyya01}.

There were researches that used SSI to guarantee application high availability ~\citep{vallee10}, to provide an information system strategy where there will be a single view of electronic mail, database access, print and plot service, and archival storage for all users ~\citep{heterick88}, to enable power-aware, adaptive, and efficient ad hoc networking applications ~\citep{rimon01}. This study proposes the use of SSI to optimize resource utilization and performance of servers even during overload conditions.

Features of SSI Clustering Systems
The major features of a single-system image cluster include the following ~\citep{buyya01}:

\begin{enumerate}
	\item Process migration: Processes migration enables processes in computer clusters to move from machine to machine, possibly for load balancing. Migration may also include other associated resources.
	\item Process checkpointing: Checkpointing allows rollback recovery after a failure by periodically saving the process state and intermediate computing results in memory or disk. Process migration and checkpointing are usually implemented together by checkpointing a process first before migrating it, then restarting it on another node.
	\item Single process space: All user processes have a unique cluster-wide process ID. A process on any node can create child processes on the same or different node and communicate with any other process on a remote node. Clusters should provide process management tools to operate on all processes in the cluster as if they are running on local machines.
	\item Single-file hierarchy: Most SSI systems provide a single view of the file system. This enables processes to run on any available node and access needed files safely.
	\item Single I/O space (SIOS): Some SSI systems allow all nodes to access disks, network-attached RAIDs and peripheral devices of other nodes. There may be some restrictions on the kinds of accesses allowed.
	\item Single entry point: Some SSI systems provide a single address visible from outside the cluster that can be used to contact the cluster as if it were one machine. The system transparently distributes the user's connection requests to different physical hosts to balance the load.
\end{enumerate}

This paper considered Kerrighed, OpenSSI and openMosix as SSI clustering systems since these systems does not only support the features needed to create a web-grid but these systems already have stable versions which can be used for research centers.

\subsection{Overview of Kerrighed}

Kerrighed is a Single System Image operating system for clusters. It added services to the traditional system such as remote paging, cooperative caching and global scheduler. It serves as a global resource manager which enhances cluster hardware use and makes access to distributed resources transparent ~\citep{vallee10}.

Kerrighed provides SSI features using a Linux container called Kerrighed container. By default on the Kerrighed system, the host system shares most of its resources with the Kerrighed container. Processes running in the Kerrighed container will have the ability to migrate from one node to another, checkpoint and restart, use distant memory, etc. ~\citep{kerrighed12}. Load balancing happens through a scheduler positioned at cluster nodes which works by acting on fork()s in a round robin fashion ~\citep{kerrighed12}. At times, when a node is under-loaded, the system detects the imbalance and migrates a process from a high-loaded node to an under-loaded node. Kerrighed also offers a configurable global process scheduler. Using the Kerrighed scheduler builder tool, dedicated scheduling policies can be easily written and hot plugged in the cluster. Any process can be migrated except processes strongly connected to a node. Processes using system V memory segment, individual and group of threads can be migrated ~\citep{lottiaux04}.

The Kerrighed migration mechanism is based on several mechanisms, such as process ghosting, containers, migrable streams and distributed file system ~\citep{lottiaux04}. 

Process ghosting is used to extract process state information and store corresponding data on a given device. This device can be a disk (process checkpointing), a network (process migration or remote process creation) or a memory (process duplication or memory checkpointing). 

The container mechanism is used to share data across nodes while ensuring data coherency. This mechanism is used to implement memory sharing, a cooperative file cache and the Kerrighed distributed file system called KerFS.

The migrable stream and mechanism is used to efficiently handle communicating process migration. Processes using pipes or sockets can be migrated with no penalty on latency or bandwidth after migration.

\subsection{Overview of OpenMosix}
	
OpenMosix was originally forked from MOSIX in 2002 ~\citep{openmosix12}. It was particularly useful for running parallel and I/O bound applications. OpenMosix development has been halted by its developers, but the LinuxPMI project is continuing development of the former OpenMosix code ~\citep{openmosix12}. The main properties of MOSIX based from ~\citep{barak95} include:

\begin{enumerate}
	\item Network transparency
	\item Dynamic load balancing - The main load-balancing algorithms are the load calculation algorithm, that measures the local load; the information dissemination algorithm and the migration consideration algorithm, that makes the final decision based on the available load information, the relative speed of the nodes and other parameters.
	\item Preemptive process migration
	\item Decentralized control and symmetry
	\item File system
\end{enumerate}

\subsection{Overview of OpenSSI}

OpenSSI was released in 2001 ~\citep{lottiaux04}. It was based on the Linux operating system, itself based on Locus. OpenSSI intends to give a platform that integrates open source cluster technologies. Currently, open source systems disk management systems such as GFS, OpenGFS, Lustre, OCFS, DRBD, a distributed lock mechanism (OpenDLM), and a load levelling derived from Mosix are integrated in OpenSSI.

For efficiency, OpenSSI migrates processes directly through the network. In such a system, the process is extracted from a node, directly sent through the network to a remote node, and a new running process is created ~\citep{vallee10}.

Migration can be done manually or automatically. Processes can be manually migrated, either by the process calling the special OpenSSI migrate(2) system call, or by writing a node number to a special file in the processes /proc directory. Processes may be automatically migrated in order to balance load across the cluster. OpenSSI uses an algorithm developed by the MOSIX project for determining the load on each node. Any process can be migrated except for processes strongly connected to a node (direct access to video or network card memory for instance). Processes using system V memory segment and group of threads can be migrated. However, individual threads cannot be migrated ~\citep{vallee10}.

OpenSSI uses Linux Virtual Server (LVS) to provide fault-tolerant load balanced IP services. Inbound network connections are received by a director node which redirects them to the least loaded server node. A node may be both a director and server. In the event of director node failure another director node takes over and the system continues to accept inbound connections ~\citep{vallee10}.

The OpenSSI software is available for various Linux distributions. The OpenSSI kernel is distribution independent but various distribution specific Linux user level systems need to be modified, for example the init process and the system startup scripts. Currently the supported distributions are Fedora Core 3 and Debian Sarge. Work is in progress to port OpenSSI to Debian Etch and Lenny ~\citep{vallee10}.

\section{Methodology}
Like the architecture in ~\citet{derobles11}, the Web architecture used in this study is presented in which each node is hosting a complete replica of the application. 

\subsubsection{HTTP Servers}

The HTTP servers are the desktop PCs along with the dedicated HTTP server. These servers will provide services to clients. Each will be configured to provide two different types of services, simple and complex. A simple service performs simple tasks for the clients and does not require long and complex database queries. A complex service performs complex tasks for the clients and usually requires more computational tasks and complex database joins. For this experiment, a complex task requires about 100 more server resources than a simple task.
To run the HTTP servers, Apache will be installed to the desktop PCs. Apache is an open source HTTP server for modern operating systems. Java Servlets will then be installed inside the Apache web server. Java Servlets provide a powerful mechanism for developing the server side components of web application.

\subsubsection{Installing and Configuring Kerrighed}

An NFS server will be used to export the Kerrighed environment. Thus, client nodes do not need to have any hard disk devices attached to take part in a cluster session since all kinds of data needed at runtime will be provided by NFS. A NFS serves a local '/' directory that can be mounted by the remote (NFS) clients. The NFS server must NOT be part of the cluster. Thus it does not need to run a Kerrighed kernel. 

Pre-requisites on the nodes

Nodes must be able to boot with PXE. PXE support is usually implemented in the NIC BIOS, and must be enabled in the BIOS setup. If the nodes cannot use PXE (by a limitation of the NIC or the BIOS), it is also possible to use one of the boot loaders available in the project Etherboot/gPXE (http://etherboot.org/wiki/download). 

Kerrighed Live CD

A Live CD based on Kerrighed 2.3.0. was downloaded. New versions of Kerrighed were already released however, they only support x86-64 architecture. Currently, the newest version of Kerrighed is 3.0.0. 

After booting the live CD, the head node will be configured to act as HTTP Server by installing Apache Tomcat, Java, and MySQL and making it host the simple and complex application described above.

Then, the nodes were added to the cluster by booting them up through PXE. Each node should be presented with a login prompt similar to the head node. After adding all the nodes to the cluster, enter the command \emph{demokrg} to any of the nodes to start the cluster.

\subsubsection{Installing and Configuring OpenMosix}

OpenMosix is made up of a kernel patch and some user-space tools. The kernel patch is needed to make the kernel capable of talking to other OpenMosix-enabled machines on the network. The user-space tools are needed in order to make an effective use of an OpenMosix-enabled kernel.
 
Linux Live CDs with OpenMosix include CHAOS, ClusterKnoppix, Dyne:bolic, and Quantian. CHAOS is a very small boot CD but it is typically not deployed on its own; cluster builders will use feature rich Linux distributions (such as Quantian or ClusterKnoppix) as a head node in a cluster to provide their application software, while the CHAOS distribution runs on the other nodes to provide power to the cluster. ClusterKnoppix is a specialized Linux distribution based on the Knoppix distribution, but which uses the OpenMosix kernel. The distribution contains an auto configuration system where new ClusterKnoppix-running computers attached to the network automatically join the cluster. Dyne:bolic is based on the Linux kernel with a focus on multimedia production, and is distributed with a large assortment of applications for audio and video manipulation. Quantian OS is a remastering of Knoppix/Debian for computational sciences and incorporates ClusterKnoppix. 

ClusterKnoppix 3.6 Live CD

ClusterKnoppix automatically makes the head node part of the cluster once booted. Thus, it is important that it received an IP address from a DHCP server. Also, the other nodes that will be added to the cluster will automatically be on the same subnet as in the head node. After the head node has already booted, it was configured to act as HTTP server by installing Apache Tomcat since ClusterKnoppix automatically installs Java and MySQL.

The following steps were done to run MySQL on ClusterKnoppix.
\begin{itemize}
	\item Create a \emph{mysqld} directory under the \emph{/var/run} directory by executing the command \emph{mkdir /var/run/mysqld}.
	\item Create a file \emph{mysqld.sock} inside the \emph{mysqld} directory just created by executing the command \emph{touch /var/run/mysqld/mysqld.sock}.
	\item Change the owner of \emph{/var/run/mysqld} to \emph{mysql} by executing the command \emph{chown -R mysql /var/run/mysqld}.
	\item Modify \emph{my.cnf}, the MySQL configuration file, located in the \emph{/etc/mysql} directory by commenting the line \emph{skip-networking}.
	\item Start MySQL by executing the command \emph{/etc /init.d/mysql start}.
	\item Connect to the database server my executing \emph{mysql -u root} then, create the required database and tables.
\end{itemize}

Then, we install a new java version since the java version automatically installed by ClusterKnoppix does not support the features that we need.  The following steps were done to install and use the new java version:
\begin{itemize}
	\item Download java 1.6 or higher, e.g., jre-6u35-linux-i586.bin.
	\item Change the permission of the file making it executable by entering the command \emph{chmod +x jre-6u35-linux-i586.bin}.
	\item Install java by executing \emph{./jre-6u35-linux-i586.bin}.
	\item To use the newer version of java, set the \emph{JAVA\_ HOME} environment variable to point to the directory where java was installed, e.g., \emph{export JAVA\_HOME = /home/knoppix/jre1.6.0\_35}.
\end{itemize}

Next, Apache Tomcat was installed and started after hosting the applications inside it.

After testing the application, a clustering-enabled terminal-server was setup by running the command \emph{knoppix-terminalopenmosixserver}.

Then, the nodes were added to the cluster by booting them up through PXE. At the boot prompt, enter \emph{knoppix 2} to put the other nodes in text mode.

\subsubsection{Installing and Configuring OpenSSI}

KNOPPIX/OpenSSI Live CD

Since OpenMosix and OpenSSI both uses Knoppix as operating system, similar steps were done to configure the head node to act as HTTP server.

On the first node (or any node already in the cluster), execute \emph{ssi-addnode}. It will ask few questions about how you want to configure your new node and they are as follows.
\begin{itemize}
	\item Enter a unique node number between 2 and 125.
	\item Enter MAC address of the new node to be added in the cluster.
	\item Enter a static IP address for the NIC. It must be unique and must be on the 10.0.0.0 network.
	\item Select (P)XE or (E)therboot as the network boot protocol for this node.
	\item Enter a node name. It should be unique in the cluster. 
	\item Save the configuration.
\end{itemize}

The program will now do all the work to admit the node into the cluster. Wait for the new node to join. A node up message on the first node's console will indicate this. If the new node hung searching for the DHCP server, try manually restarting the DHCP server on the head node by executing the command \emph{invoke-rc.d dhcp restart}. Also, it has been observed that sometimes the TFTP server will not respond to a client more than once. To solve this, restart inetd on the head node if client could get IP address, but could not continue booting by executing the command invoke-rc.d inetd restart. You can confirm its membership with the cluster command \emph{cluster \---v}.

\subsubsection{Performance Evaluation}
Next, we showed that these architectures were able to optimize the performance of the server as characterized by the average total access delays perceived by clients. This was done through an online technique where a program was created to generate requests of increasing workloads. These requests are fed to the to the architectures configured with Kerrighed, openMosix, and OpenSSI. We calculated the total response time of the different SSI clustering systems and compared their performance. The total response time was computed by calculating the time elapsed from sending the request to receiving the response from the server which includes the traffic, load-balancing overhead, and the server processing time. These are recorded using a functionality included in the program created for generating client requests. The information gathered are stored in a text file and was later used as the basis for analyzing the total response time.

We evaluated the performance of the web-grid by subjecting it to varying workloads, characterized by the number of requests it received per second. Here, we looked at workloads with values 2, 5, 10, 15, 20, 25, 30, 35, 40, 45, 50, 100, 150, 300, 350, 400, 40, 500, and 1000 requests per second (rps). For each workload, we observed the effect of increasing the number of PCs n involved in the grid on total response time. We used $n$ = 2, 4, and 6 PCs.

We wish to measure that the response time $t$ is a function of the factorial effect of three factors namely: 1)~SSI clustering system ${x_1}$ with three discrete levels, 2)~workload ${x_2}$ with 21 continuous levels; and 3)~web-grid size ${x_3}$ with three continuous levels. Each ${x_1}\times{x_2}\times{x_3}$ combination was replicated five times. The model of which is shown in Equation~\ref{eqn:model1}.

\begin{eqnarray}
	t &=& \alpha_1x_1+\alpha_2x_2+\alpha_3x_3+\alpha_4x_1x_2+\alpha_5x_1x_3+\nonumber\\
            & & \quad\alpha_6x_2x_3+\alpha_7x_1x_2x_3+\epsilon_{123} \label{eqn:model1}
\end{eqnarray}

where $\alpha$s are coefficients and $\epsilon_{123}$ is the residual effect brought about by random error.

We hypothesize that the coefficient $\alpha_7$ is non-zero which means that three-way interaction is present. In the event that $\alpha_7$ is not significantly different from zero at 5\% confidence level, we hypothesize that either $\alpha_4$, $\alpha_5$, or $\alpha_6$ is non-zero at the same confidence level which means that two-way interaction is present. Otherwise, we hypothesize that either $\alpha_1$, $\alpha_2$, or $\alpha_3$ is not zero also at the same confidence level which means that the factors we considered have no interaction.

We also wish to measure that the number of misses $m$ as a function of the factorial effect of the same three factors. The model of which is shown in Equation~\ref{eqn:model2}.

\begin{eqnarray}
	m &=& \beta_1x_1+\beta_2x_2+\beta_3x_3+\beta_4x_1x_2+\beta_5x_1x_3+\nonumber\\
            & & \quad\beta_6x_2x_3+\beta_7x_1x_2x_3+\epsilon_{123} \label{eqn:model2}
\end{eqnarray}

where $\beta$s are coefficients and $\epsilon_{123}$ is the residual effect brought about by random error.

We hypothesize that the coefficient $\beta_7$ is non-zero which means that three-way interaction is present. In the event that $\beta_7$ is not significantly different from zero at 5\% confidence level, we hypothesize that either $\beta_4$, $\beta_5$, or $\beta_6$ is non-zero at the same confidence level which means that two-way interaction is present. Otherwise, we hypothesize that either $\beta_1$, $\beta_2$, or $\beta_3$ is not zero also at the same confidence level which means that the factors we considered have no interaction.

To measure the respective hypotheses for $\alpha$ and $\beta$, we conducted analysis of variance for $t$ and $m$ using the models in Equations~\ref{eqn:model1} and ~\ref{eqn:model2}.

\section{Results and Discussion}
Using our experimental testbed, the performance of the three SSI clustering systems were evaluated.

\subsection{Response Time}
\begin{figure*}[htb]
        \centering
        \epsfig{file=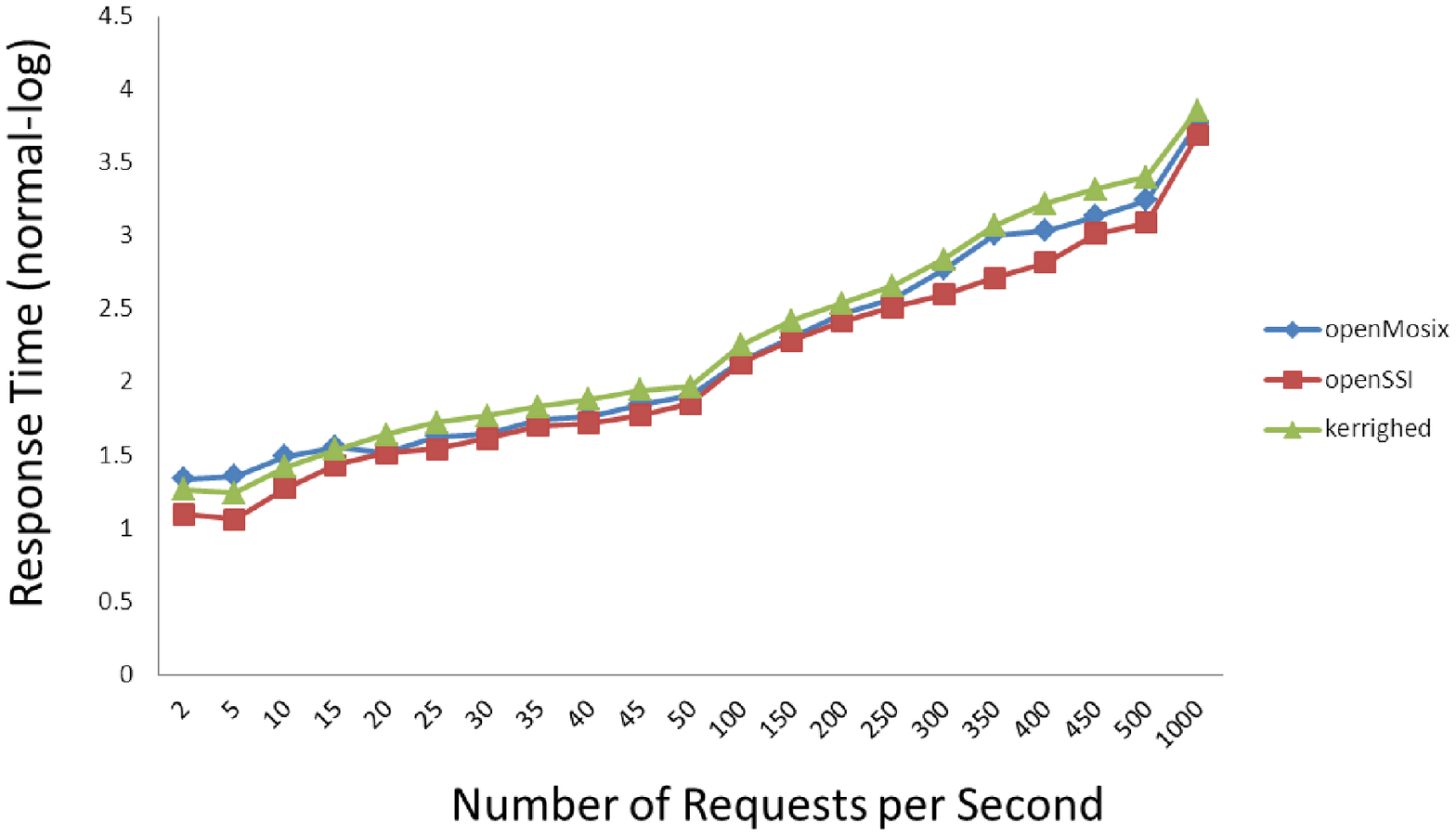, width=6in}
        \caption{The normal-log plot of $t$ when the cluster is doing simple jobs as $x_2$ is increased for each $x_1$ averaged accross $x_3$ and $R$.}
        \label{res1}
\end{figure*}

Table~\ref{table1} shows the analysis of variance table for $t$ when the cluster is doing simple jobs. From the table, we will see that the $R$ is significant at 0.1\% confidence level. This means that we were successful in replicating each ${x_1}\times{x_2}\times{x_3}$ combination. We also see from the table that the combination ${x_1}\times{x_2}$ is also siginificant at 0.1\% level. Figure~\ref{res1} shows the normal-log plot of $t$ as $x_2$ is increased for each $x_1$ averaged accross $x_3$ and $R$. This plot shows the cause of interaction between $x_1$ and $x_2$.

\begin{table*}
	\caption{Analysis of Variance table for $t$ for simple jobs. $R$ stands for replicates, Df stands for Degrees of Freedom, Sum Sq stands for Sum of Squares, Mean Sq stands for Mean Square and Pr stands for Probability. * means siginificant at 5\% confidence level, ** means siginificant at 1\% confidence level and *** means significant at 0.1\% confidence level.}
	\label{table1}
	\centering
	\begin{tabular}{l|rrrrr}
		\hline\hline
		Variation&Df&Sum Sq&Mean Sq&F value&Pr(>F)\\
		\hline
		$R$&4&1361074.48&340268.62&9.36&0***\\
		$x_1$&2&18025700.16&9012850.08&247.94&0***\\
		$x_2$&20&1593014626.00&79650731.28&2191.20&0***\\
		$x_3$&2&198097.95&99048.97&2.72&0.07\\
		$x_1 \times x_2$&40&55318920.38&1382973.01&38.05&0***\\
		$x_1 \times x_3$&4&292937.84&73234.46&2.01&0.09\\
		$x_2 \times x_3$&40&877512.22&21937.81&0.60&0.98\\
		$x_1 \times x_2 \times x_3$&80&1354073.65&16925.92&0.47&1.00\\
		\hline
		Total&752&27335456.63&36350.34\\
		\hline\hline
	\end{tabular}
\end{table*}

\begin{figure*}[htb]
        \centering
        \epsfig{file=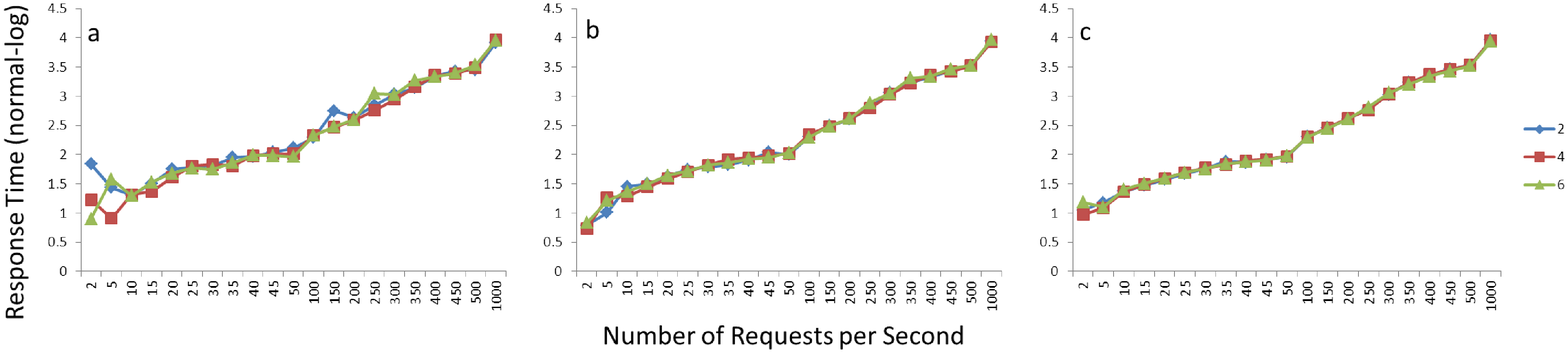, width=6in}
        \caption{The normal-log plot of $t$ when the cluster is doing complex jobs at each $x_1$ level respectivley as $x_2$ is increased for each $x_3$ averaged accross $R$.}
        \label{res2}
\end{figure*}

Table~\ref{table2} shows the analysis of variance table for $t$ when the cluster is doing complex jobs. From the table, we will see that the three-factor combination $x_1 \times x_2 \times x_3$ is siginificant at 0.1\% level. Figure~\ref{res2} a,b,c show the normal-log plot of $t$ at each $x_1$ level respectivley as $x_2$ is increased for each $x_3$ averaged accross $R$. This plot shows the cause of interaction between $x_1$, $x_2$ and $x_3$.

\begin{table*}
	\caption{Analysis of Variance table for $t$ for complex jobs. $R$ stands for replicates, Df stands for Degrees of Freedom, Sum Sq stands for Sum of Squares, Mean Sq stands for Mean Square and Pr stands for Probability. * means siginificant at 5\% confidence level, ** means siginificant at 1\% confidence level and *** means significant at 0.1\% confidence level.}
	\label{table2}
	\centering
	\begin{tabular}{l|rrrrr}
		\hline\hline
		Variation&Df&Sum Sq&Mean Sq&F value&Pr(>F)\\
		\hline
		$R$&4&135220.15&33805.04&0.89&0.47\\
		$x_1$&2&117315.29&58657.64&1.55&0.21\\
		$x_2$&20&3741202837.51&187060141.88&4943.67&0***\\
		$x_3$&2&183240.52&91620.26&2.42&0.09\\
		$x_1 \times x_2$&40&2214882.24&55372.06&1.46&0.03*\\
		$x_1 \times x_3$&4&540881.64&135220.41&3.57&0.01**\\
		$x_2 \times x_3$&40&2086187.19&52154.68&1.38&0.06\\
		$x_1 \times x_2 \times x_3$&80&6259327.49&78241.59&2.07&0***\\
		\hline
		Total&752&28454423.10&37838.33\\
		\hline\hline
	\end{tabular}
\end{table*}

\subsection{Missed Requests}
\begin{figure*}[htb]
        \centering
        \epsfig{file=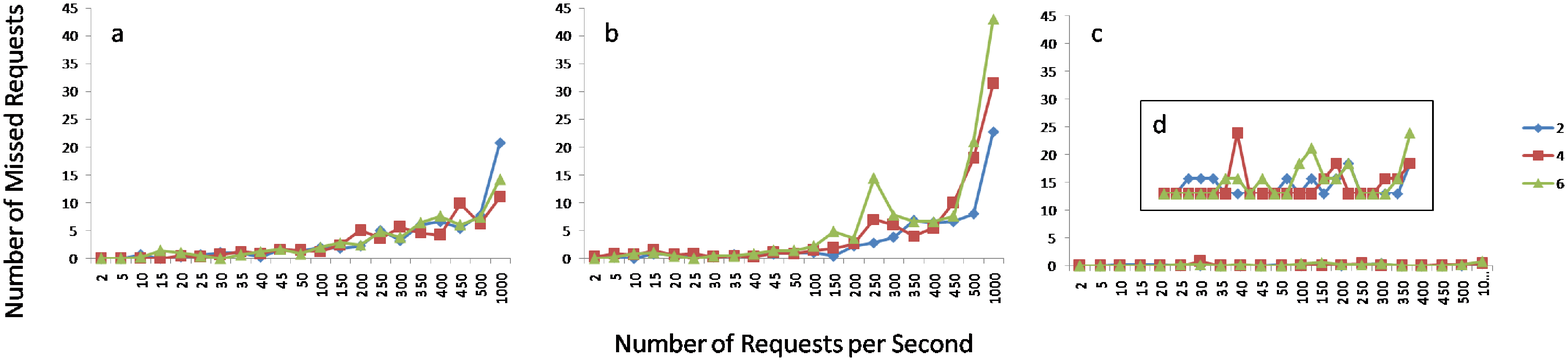, width=6in}
        \caption{The normal-log plot of $m$ when the cluster is doing simple jobs at each $x_1$ level respectivley as $x_2$ is increased for each $x_3$ averaged accross $R$.}
        \label{res3}
\end{figure*}

Table~\ref{table3} shows the analysis of variance table for $m$ when the cluster is doing simple jobs. From the table, we will see that the three-factor combination $x_1 \times x_2 \times x_3$ is siginificant at 0.1\% level. Figure~\ref{res3} a,b,c show the normal-log plot of $t$ at each $x_1$ level respectivley as $x_2$ is increased for each $x_3$ averaged accross $R$. This plot shows the cause of interaction between $x_1$, $x_2$ and $x_3$.

\begin{table*}
	\caption{Analysis of Variance table for $m$ for simple jobs. $R$ stands for replicates, Df stands for Degrees of Freedom, Sum Sq stands for Sum of Squares, Mean Sq stands for Mean Square and Pr stands for Probability. * means siginificant at 5\% confidence level, ** means siginificant at 1\% confidence level and *** means significant at 0.1\% confidence level.}
	\label{table3}
	\centering
	\begin{tabular}{l|rrrrr}
		\hline\hline
		Variation&Df&Sum Sq&Mean Sq&F value&Pr(>F)\\
		\hline
		$R$&4&123.97&30.99&4.15&0**\\
		$x_1$&2&3162.18&1581.09&211.59&0***\\
		$x_2$&20&12473.87&623.69&83.47&0***\\
		$x_3$&2&132.58&66.29&8.87&0***\\
		$x_1 \times x_2$&40&8466.44&211.66&28.33&0***\\
		$x_1 \times x_3$&4&280.48&70.12&9.38&0***\\
		$x_2 \times x_3$&40&534.04&13.35&1.79&0**\\
		$x_1 \times x_2 \times x_3$&80&1481.83&18.52&2.48&0***\\
		\hline
		Total&752&5619.23&7.47\\
		\hline\hline
	\end{tabular}
\end{table*}

\begin{figure*}[htb]
        \centering
        \epsfig{file=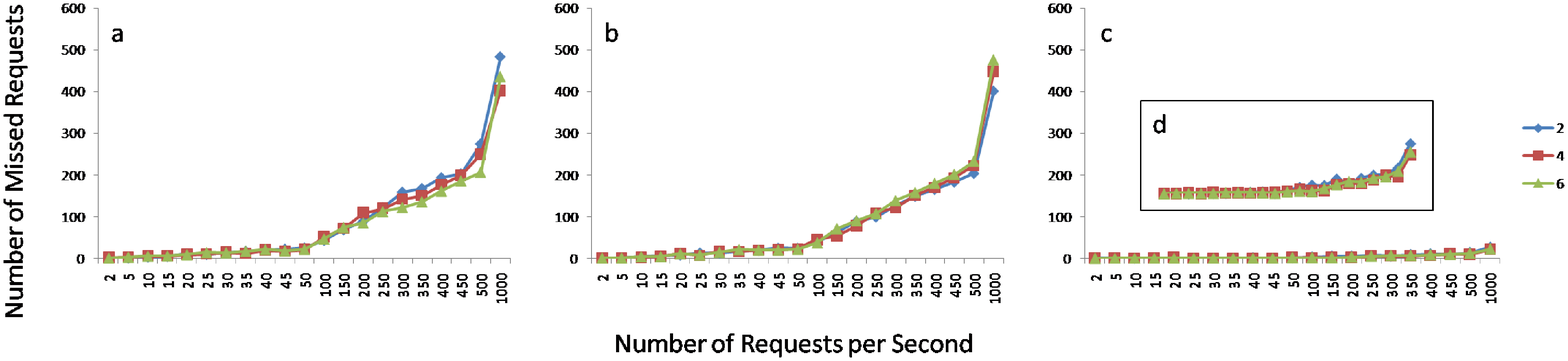, width=6in}
        \caption{The normal-log plot of $m$ when the cluster is doing complex jobs at each $x_1$ level respectivley as $x_2$ is increased for each $x_3$ averaged accross $R$.}
        \label{res4}
\end{figure*}

Table~\ref{table4} shows the analysis of variance table for $m$ when the cluster is doing complex jobs. From the table, we will see that the three-factor combination $x_1 \times x_2 \times x_3$ is siginificant at 0.1\% level. Figure~\ref{res4} a,b,c show the normal-log plot of $t$ at each $x_1$ level respectivley as $x_2$ is increased for each $x_3$ averaged accross $R$. This plot shows the cause of interaction between $x_1$, $x_2$ and $x_3$.

\begin{table*}
	\caption{Analysis of Variance table for $m$ for complex jobs. $R$ stands for replicates, Df stands for Degrees of Freedom, Sum Sq stands for Sum of Squares, Mean Sq stands for Mean Square and Pr stands for Probability. * means siginificant at 5\% confidence level, ** means siginificant at 1\% confidence level and *** means significant at 0.1\% confidence level.}
	\label{table4}
	\centering
	\begin{tabular}{l|rrrrr}
		\hline\hline
		Variation&Df&Sum Sq&Mean Sq&F value&Pr(>F)\\
		\hline
		$R$&4&2784.04&696.01&4.01&0***\\
		$x_1$&2&1347391.25&673695.62&3877.80&0***\\
		$x_2$&20&5021564.55&251078.23&1445.21&0***\\
		$x_3$&2&431.84&215.92&1.24&0.29\\
		$x_1 \times x_2$&40&2156918.31&53922.96&310.38&0***\\
		$x_1 \times x_3$&4&9619.40&2404.85&13.84&0\\
		$x_2 \times x_3$&40&6777.45&169.44&0.98&0.52\\
		$x_1 \times x_2 \times x_3$&80&43924.24&549.05&3.16&0\\
		\hline
		Total&752&130645.96&173.73\\
		\hline\hline
	\end{tabular}
\end{table*}

In terms of $m$, Kerrighed has the least missed requests while its $t$ is comparable with the other two. Thus, we recommend the use of the system to web-grids.

\section{Summary and Conclusion}
A prototype distributed desktop web-grid was developed to improve the performance of both static and dynamic-content Web sites, characterized by client access time and resource utilization, even during overload conditions. Through an on-line simulation technique, performance savings achieved by the proposed algorithm was explored. Three different SSI clustering systems were evaluated: openMosix, OpenSSI and Kerrighed.

With the new architecture, utilization of resources, especially servers or computers were optimized. Allocation of additional resources to handle the peak workload caused by Time-of-Day effects and Flash crowd arrivals are no longer necessary. Instead, computers such as those belonging to secretaries, which are used only for simple tasks such as typing, replaced the task of the servers in providing services to clients. Thus, a significant power and other resources savings was achieved.

Results showed that for $t$, the combination ${x_1}\times{x_2}$ is siginificant at 0.1\% level when the cluster is doing simple jobs.  Results also showed that for $t$, the three-factor combination $x_1 \times x_2 \times x_3$ is siginificant at 0.1\% level when the cluster is doing complex jobs. Also, the analysis of variance for $m$ for both simple and complex jobs showed that the three-factor combination $x_1 \times x_2 \times x_3$ is siginificant at 0.1\% confidence level.

Kerrighed can be considered as the best SSI clustering system because it has the least $m$ while $t$ is comparable with others.

\bibliographystyle{plainnat}
\bibliography{web-grid}


\end{document}